\documentclass[twocolumn,epjc3]{svjour3}          

\RequirePackage[T1]{fontenc}

\smartqed  

\RequirePackage{graphicx}
\RequirePackage{mathptmx}      
\RequirePackage{flushend}
\RequirePackage[numbers,sort&compress]{natbib}
\RequirePackage[colorlinks,citecolor=blue,urlcolor=blue,linkcolor=blue]{hyperref}
\RequirePackage{amssymb,amsmath}

\begin{document}

\title{Probing gravitational non-minimal coupling with dark energy surveys}

\author{
Chao-Qiang Geng\thanksref{e1,addr1,addr2,addr3} \and
Chung-Chi Lee\thanksref{e2,addr3} \and
Yi-Peng Wu\thanksref{e3,addr4}
}

\thankstext[$\star$]{t1}{Thanks to the title}
\thankstext{e1}{e-mail: geng@phys.nthu.edu.tw}
\thankstext{e2}{e-mail: g9522545@oz.nthu.edu.tw}
\thankstext{e3}{e-mail: ypwu@phys.sinica.edu.tw}

\institute{Chongqing University of Posts \& Telecommunications, 
Chongqing, 400065, China\label{addr1}
\and
Department of Physics, National Tsing Hua University, Hsinchu 30013, Taiwan\label{addr2}
\and
National Center for Theoretical Sciences, Hsinchu 30013, Taiwan\label{addr3}
\and
Institute of Physics, Academia Sinica, Taipei 11529, Taiwan\label{addr4}
}

\date{Received: date / Accepted: date}

\maketitle

\begin{abstract}
We investigate observational constraints on a specific one-parameter extension to the minimal quintessence model, where the quintessence field acquires a quadratic coupling to the scalar curvature through a coupling constant $\xi$. 
The value of $\xi$ is highly suppressed in typical tracker models if the late-time cosmic acceleration is driven at some field values near the Planck scale.
We test $\xi$ in a second class of models in which the field value today becomes a free model parameter. We use the combined data from type-Ia supernovae, cosmic microwave background, baryon acoustic oscillations and matter power spectrum, to weak lensing measurements and find a best-fit value $\xi > 0.289$ where $\xi = 0$ is excluded outside the 95 per cent confidence region. 
The effective gravitational constant $G_{\rm eff}$ subject to the hint of a non-zero $\xi$ is constrained to $-0.003 < 1- G_{\rm eff}/G < 0.033$ at the same confidence level on cosmological scales, and can be narrowed down to $1- G_{\rm eff}/G < 2.2 \times 10^{-5}$ when combining with Solar System tests. 
\end{abstract}

\section{Introduction}

Cosmological observations in support of the late-time cosmic acceleration, such as measurements for the Type-Ia supernovae~\cite{Riess:1998cb,Perlmutter:1998np}, the cosmic microwave background (CMB) radiation~\cite{Ade:2013zuv,Ade:2015xua}, the large-scale structure and baryon acoustic oscillations \cite{Anderson:2012sa,Anderson:2013zyy}, strongly suggest that the current Universe density must consist of some unfamiliar ``negative pressure matter,'' namely, dark energy. Dark energy can be explained by a cosmological constant ($\Lambda$) with an extremely tiny density $\rho_{\Lambda}\approx 10^{-47}$ $\mathrm{GeV}^4$, comparing with the typical energy scales in particle physics. Another possible candidate for such a source, as legitimate as imposing $\Lambda$, is referred to a quintessence component whose dynamics is described by a scalar field $\phi$~\cite{Caldwell:1997ii}.

Quintessence with sufficiently flat potentials exhibits attractor solutions such that wide ranges of initial conditions approach the scalar field dominated universe ($\Omega_\phi=1$). One of the common forms  is the exponential potential $V(\phi)\sim e^{-\lambda\kappa\phi}$ with $\kappa^2 = 8 \pi G$, where the cosmic acceleration is realized if $\lambda^2 < 2$ \cite{Ferreira:1997au, Ferreira:1997hj, Copeland:1997et, Billyard:1998hv}. These late-time attractor solutions, however, demand nearly fixed values of the quintessence density $\rho_\phi$ all along the evolution history of the universe. The fact that $\rho_\phi$ coincides with the (dust-like) matter density today implies that the initial densities of the two species have to be in some huge yet precise hierarchy, leading to the cosmic coincidence problem \cite{Zlatev:1998tr}.

On the other hand, tracker fields are introduced as a specific class of quintessence aiming to solve the cosmic coincidence problem \cite{Zlatev:1998tr, Steinhardt:1999nw}. The main advantage of a tracker field is that the initial value of $\Omega_\phi$ can be close to unity to within few orders of magnitude, and thus it accepts the assumption of equipartition after inflation \cite{Zlatev:1998tr, Steinhardt:1999nw}. Typical tracker solutions have been found in potentials of the inverse power-law $V(\phi)\sim\phi^{-\alpha}$ with $\alpha > 0$ \cite{Peebles:1987ek, Ratra:1987rm, Ferreira:1997au}. Unfortunately, recent observational constraints on the tracker fields require $\alpha\ll 1$ \cite{Chiba:2009gg, Wang:2011bi,Chiba:2012cb, Tsujikawa:2013fta}, which significantly narrows down the tracking basin of attraction \cite{Bludman:2004az}. To mimic a cosmological constant by taking $\alpha\rightarrow 0$, the present energy density must be realized from the fine-tuned initial condition $\rho_\phi \simeq 10^{-47}$ $\mathrm{GeV}^4$, only to duplicate the cosmic coincidence problem.

Given these considerations, much effort has been devoted to the extended scenario in which the quintessence field is non-minimally coupled to the scalar curvature $R$ (\cite{Perrotta:1999am, Baccigalupi:2000je}, see also \cite{deRitis:1999zn, Amendola:1999qq, Uzan:1999ch, Chiba:1999wt, Bartolo:1999sq, Faraoni:2000wk, Carvalho:2004ty, Sen:2009mc, Gupta:2009kk}) through a non-minimal coupling $\xi R \phi^2$, where $\xi$ is the coupling constant.
There exist tracking solutions purely induced by the non-minimal coupling such that a very wide range of initial values of $\rho_\phi$ can evolve toward the late-time attractor of the cosmic acceleration, even if the potential is extremely flat \cite{Matarrese:2004xa}. This gravitational effect is manifested under conditions of the R-boost \cite{Baccigalupi:2000je} or the slow-roll \cite{Chiba:2010cy}. In either case, the gravitational tracker solutions are obtained without assuming particular forms of the potentials.

A fine-tuned potential, however, is needed to achieve a preferable cosmic acceleration, despite that the gravitational tracker solutions help to relax the cosmic coincidence problem. In particular, when the potential is exactly flat the extended quintessence model \cite{Perrotta:1999am} can be cast into the prototype Brans-Dicke (PBD) theory \cite{Brans:1961sx, Clifton:2011jh} but with a time-varying Brans-Dicke parameter $\omega$ depending on the value of $\phi$. Constraints on the value of $\omega$ have been widely studied from the CMB anisotropy and structure formations \cite{Nagata:2003qn, Acquaviva:2004ti, Wu:2009zb, Li:2013nwa, Avilez:2013dxa}, the parametrized post-Newtonian parameters \cite{Bertotti:2003rm, Williams:2004qba, Williams:2004uw, Perivolaropoulos:2009ak}, and the big-bang nucleosynthesis \cite{Clifton:2005xr, Coc:2006rt, Coc:2008yu}. These constraints in terms put a tight bound on the non-minimal coupling as $\vert\xi\vert < 10^{-2}$ in the inverse power-law model of the \cite{Ratra:1987rm} type, given that the scalar field must reach a Planck scale by now ($\phi_0 \sim 10^{19}$ GeV) after rolling on the track from arbitrary initial conditions \cite{Baccigalupi:2000je}.

In this work, we show that the restriction on $\xi$ shall be significantly reduced in other types of potentials where the present day value of field can be freely determined by its initial conditions. As a concrete example, we put the quintessence field in the slow-roll exponential potential type \cite{Ferreira:1997au, Ferreira:1997hj, Copeland:1997et, Billyard:1998hv} where the field value is nearly frozen all along the expansion history when $\lambda\rightarrow 0$. We investigate the effects of $\xi$ from an analytic approach to the background equation of state $w$ and to the effective gravitational constants on cosmological scales or inside the Solar System. These modified gravitational constants are sketched by the phenomenological functions $\mu$ and $\Sigma$, which can be useful for testing the deviation from the $\Lambda$CDM model \cite{Pogosian:2010tj, Hojjati:2011xd, Zuntz:2011aq}. 
We then confront the exponential model with the recent observational data to fit the most preferable value of $\xi$ on cosmological scales. We pick up the absolutely flat model ($\lambda  = 0 $) as our paradigm for which its background dynamics is identical to that of $\Lambda$CDM where $\xi$ only modifies the growth dynamics through $\mu$ or $\Sigma$. It is also interesting that the flat model with a constant potential has only the same free parameters as that of the PBD theory. Finally, we study the possible near future cosmic evolution based on a non-zero $\xi$.

This article is organized as follows: In Sec.~\ref{sec:2}, we review basic equations of a non-minimally coupled scalar field. In Sec.~\ref{sec:3}, we provide the analytic and numerical results of the field evolution in the exponential potential model, including the constant potential model as a special case. We use observational data to obtain the best-fit or constraints on the model parameters, as shown in Sec.~\ref{sec:4}. We present summary and discussions in Sec.~\ref{sec:5}. 

\section{Extended quintessence models } \label{sec:2} 

\subsection{Background equations}
In this work we shall focus on the one-parameter extension of the minimal dark energy scenario~\cite{Caldwell:1997ii} that the quintessence field $\phi$ has an explicit coupling to the scalar curvature $R$ given by the action
\begin{equation}
\mathcal{S} = \mathcal{S}_\phi + \mathcal{S}_M,
\end{equation}
where $\mathcal{S}_M$ is the action for both the relativistic and nonrelativistic matter, and
\begin{equation}
\label{eq:action_STmodel}
\mathcal{S}_\phi = \int d^4x \sqrt{-g} \left[  \left( 1 + \xi \kappa^2 \phi^2 \right) \frac{R}{2\kappa^2} - \frac{1}{2} \left( \nabla \phi \right)^2 - V(\phi) \right]\,,
\end{equation}
with the constant $\xi$ representing the non-minimal coupling parameter which exhibits the minimal value at $\xi=0$ and the conformal one at $\xi = -1/6$. 
If the potential is simply a constant, say $V(\phi) = \Lambda$,
the model \eqref{eq:action_STmodel} can be cast into the PBD
theory \cite{Brans:1961sx, Clifton:2011jh} through the field redefinition 
$\Phi = (1+\xi\kappa^2\phi^2)/2$, leading to
\begin{equation}
\label{eq:action_BDmodel}
\mathcal{S}_\phi = \mathcal{S}_\Phi \equiv
\frac{1}{\kappa^2} \int d^4x \sqrt{-g} \left[ \Phi R - \frac{\omega}{\Phi} \left( \nabla \Phi \right)^2 - \kappa^2\Lambda \right]  ,
\end{equation}
with the time evolving Brans-Dicke parameter 
\begin{equation}
\label{eq:omega}
\omega(\phi) = \frac{1+\xi\kappa^2\phi^2}{4\xi^2 \kappa^2\phi^2}.
\end{equation}
Therefore a massless extended quintessence is a special scalar-tensor theory that has only the same number of model parameters as that of the PBD theory.
Note that the positive energy condition $1+\xi\kappa^2\phi^2 > 0$ 
guarantees $\omega > 0$ so that the model \eqref{eq:action_STmodel} can satisfy the no-ghost condition $\omega > -3/2$ in the Brans-Dicke theory \cite{Clifton:2011jh}.
The Solar System measurements report a very strong constraint 
$\omega > 40000$ on the Brans-Dicke parameter (\cite{Bertotti:2003rm, Williams:2004qba, Williams:2004uw}), 
but this constraint is significantly reduced to $\omega > 692$ on cosmological scales \cite{Avilez:2013dxa}. 

The variation of the action~\eqref{eq:action_STmodel} with respect to the metric $g_{\mu \nu}$ results in the modified Einstein field equation
\begin{equation}\label{eq:field_STmodel}
\left( 1 + \xi \kappa^2 \phi^2 \right) G_{\mu\nu}
= \kappa^2 \left( T_{\mu\nu} + \Theta_{\mu\nu}\right),
\end{equation}
where
\begin{equation}
\Theta_{\mu\nu} =
\nabla_{\mu}\phi \nabla_{\nu}\phi - 
g_{\mu\nu}\left[ \frac{( \nabla\phi )^2}{2}   +  V \right] +
\xi \left(\nabla_{\mu} \nabla_{\nu} -g_{\mu\nu} \nabla^2 \right) \phi^2.
\end{equation}
Here $G_{\mu\nu}$ is the Einstein tensor and $T_{\mu\nu}$ is the energy-momentum tensor of matter.
Taking the spatially flat Friedmann-Robertson-Walker (FRW) background
\begin{equation}
\label{eq:FRWmetric}
ds^2=-dt^2+a^2(t) d \vec{x}^{\,2} \,,
\end{equation}
with the scale factor $a(t)$, the background field equations are given by
\begin{eqnarray}
& H^2=\frac{\kappa^2}{3}\left( \rho_M + \rho_{\phi} \right) \,, \label{eq:eom_ST_1} \\
& \dot{H}=-\frac{\kappa^2}{2} 
   \left( \rho_M + \rho_{\phi} +P_M +P_{\phi} \right) \,, \label{eq:eom_ST_2} \\
&\rho_{\phi} = \frac{\dot{\phi}^2}{2}+V-3\xi H^2 \phi^2 -6\xi H \phi \dot{\phi} \,, \label{eq:rho_phi_ST} \\
&P_{\phi} = \frac{\dot{\phi}^2}{2} -V 
+ \xi \left( 2\dot{H} +3 H^2 \right) \phi^2 +2 \xi \left( \ddot{\phi} \phi 
+\dot{\phi}^2 +2H \phi \dot{\phi} \right), \label{eq:P_phi_ST}
\end{eqnarray}
where $\rho_M$ and $P_M$ are the energy density and pressure of matter, while $\rho_{\phi}$ and $P_{\phi}$ are the effective energy density and pressure of the scalar field, respectively.

The equation of motion for the scalar field is govern by the Klein-Gordon equation 
\begin{eqnarray}
\label{eq:Klein_Gordon}
\square \phi+\xi R \phi - V_\phi=0,
\end{eqnarray}
where $\square\equiv\nabla^\mu\nabla_\mu$ and $V_\phi\equiv dV/d\phi$.
In the homogeneous FRW background, the Klein-Gordon equation takes the form of
\begin{equation}
\label{eq:eom_ST_phi}
\ddot{\phi} + 3H \dot{\phi} + V_{\phi} - 6 \xi \left( \dot{H}+2H^2 \right)\phi = 0 \,,
\end{equation}
where we have used $R=6\dot{H} + 12 H^2$. 
The background equation~\eqref{eq:eom_ST_phi} 
can be regained from the continuity equation of $\rho_\phi$ and $P_\phi$:
\begin{equation}
\label{eq:continuity_ST}
\dot{\rho}_{\phi} + 3H \left( \rho_{\phi} + P_{\phi} \right) = 0 \,,
\end{equation}
so that the non-minimally coupled scalar field is effectively a perfect fluid with an equation of state 
$w_\phi=P_\phi/\rho_\phi$. 

In terms of the e-folding $N=\ln a$, the Klein-Gordon equation~\eqref{eq:eom_ST_phi} becomes
\begin{equation}
\label{eq:eom_efolding1}
\phi^{\prime\prime} +\left( 3 + \frac{H^\prime}{H} \right) \phi^{\prime} - 6 \xi \left( 2 + \frac{H^\prime}{H} \right) \phi + \frac{\kappa V_\phi}{H^2} = 0,
\end{equation}
where the primes are e-folding derivatives, and the Planck unit has been used for $\phi$. Taking $H^\prime = -3/2(1+w)H$, Eq.~\eqref{eq:eom_efolding1} can be rewritten as
\begin{equation}
\label{eq:eom_efolding2}
\phi^{\prime\prime} + \frac{3}{2}( 1 - w ) \phi^{\prime} - 3 \xi \left( 1 - 3 w \right) \phi + \frac{\kappa V_\phi}{H^2} = 0.
\end{equation} 
Similarly, the first Friedmann equation~\eqref{eq:eom_ST_1} can be given in terms of the density fraction as
\begin{equation}
1 = \Omega_r + \Omega_m + \Omega_\phi,
\end{equation}
where $\Omega_r$ ($\Omega_m$) is the density fraction of radiation (dust-like matter) with today's value $\Omega_{m0} = 0.308$ \cite{Ade:2015xua}, and $\Omega_\phi$ is the density of the scalar field, which is further divided into
\begin{equation}
\Omega_\phi = \frac{\kappa^2 \rho_\phi}{3H^2} = \Omega_V + \Omega^{nc}_\phi,
\end{equation}
where $\Omega_V = \kappa^2 V/ (3H^2)$ and 
\begin{equation}
\Omega^{nc}_\phi = \frac{1}{6}\phi^{\prime 2} - \xi \phi^2 -2\xi\phi\phi^\prime.
\end{equation}
Note that we have included the kinetic term $\dot\phi^2/2$ into $\Omega^{nc}_\phi$
since it depends nontrivially on the parameter $\xi$.

\subsection{Perturbation functions}
Let us characterize the effects of the non-minimal coupling to the gravitational constants in the linear density perturbations. The full perturbed Einstein equations are provided in \ref{sec:LPE} (see also \cite{Tsujikawa:2007gd} and \cite{Pettorino:2008ez}). For the study of non-linear perturbations one may refer to \cite{Pace:2013pea, Fan:2015lta} 

In a non-minimally coupled theory, the scalar field fluctuations give rise to anisotropy between the curvature perturbation $\Phi$ and the Newtonian potential $\Psi$ (as defined in Eq. \eqref{eq:newtonian_gauge}). On subhorizon scales where $k \gg aH$ and in the Newtonian limit where time derivatives are negligible with respect to spatial derivatives, the anisotropy parameter $\gamma(\phi, k)= \Phi/\Psi$ is led by
\begin{equation}
\gamma = \frac{1+ \omega( 1 + \beta)}{2 +\omega(1 + \beta)},
\end{equation}
where $\beta(k) = a^2M^2/k^2$ and $M^2 = d^2V(\phi)/d \phi^2$.

The modified Poisson equation for the dust-like matter and the relativistic matter are given by \cite{Hojjati:2011xd}
\begin{eqnarray}
\label{eq:Poisson}
& k^2\Psi = - 4 \pi G \mu(\phi, k)\, a^2 \rho_M \Delta, \\
\label{eq:Weyl}
& k^2(\Phi + \Psi) = - 8 \pi G \Sigma(\phi)\, a^2 \rho_M \Delta,
\end{eqnarray}
respectively, where $\Delta$ is the comoving matter density perturbation. In the extended quintessence model \eqref{eq:action_STmodel}, we have a scale-independent function for the lensing effect ($\phi$ in Planck unit)
\begin{equation}
\Sigma = \frac{1}{1+ \xi \phi^2},
\end{equation}
and a scale-dependent function for the matter growth
\begin{equation}
\mu = \frac{2\Sigma}{1+\gamma}
    =  \frac{4+2\omega( 1 + \beta)}{3+2\omega( 1 + \beta)} \Sigma. 
\end{equation}
It is possible to choose a present value $\phi_0 = 0$ such that $\gamma_0 = \mu_0 = \Sigma_0 = 1$, which coincide with the predictions in general relativity.
Constraints on the deviation from general relativity are $1- \mu_0 = -0.05\pm 0.25$ and $1- \Sigma_0 = 0.00\pm 0.14$ for the fiducial $\Lambda$CDM background expansion at the 68 per cent confidence level \cite{Simpson:2012ra}. Some tension with the $\Lambda$CDM prediction is reported by the combined Planck CMB polarization and low multipole data with the BAO and weak lensing measurement \cite{Ade:2015rim}.

We can define $\kappa^2_{\rm eff} = 8\pi G_{\rm eff} = 8\pi G\Sigma$ to rewrite the Einstein equation \eqref{eq:field_STmodel} as
\begin{equation}
\label{eq:Einstein_eff}
G_{\mu\nu} = \kappa_{\rm eff}^2 \left( T_{\mu\nu} + \Theta_{\mu\nu}\right),
\end{equation}
so that the Friedmann Eqs. \eqref{eq:eom_ST_1} and \eqref{eq:eom_ST_2} take the same forms as those in the minimal quintessence model upto the effective constant $G_{\rm eff}$ when $\dot\phi$ is negligible.
Let us consider the usual slow-roll potential of the exponential form
\begin{equation}
\label{eq:V}
V(\phi) = V_0 e^{-\lambda \phi},
\end{equation}
where $M^2 = \lambda^2  V$ with $\phi$ in the Planck unit. Suppose that $\lambda =0.01$ and $\phi_0 =1$, one can easily check that the energy difference of the potential $\vert V(\phi)/V(\phi_0)-1\vert \lesssim 0.01$ for $\phi \in [0,2]$.
This model recovers the PBD theory when $\lambda = 0$,
in which $\beta = M = 0$ and 
\begin{eqnarray}
& \gamma_{\mathrm{BD}} = \frac{1+ \omega}{2 +\omega}, \\
& \mu_{\mathrm{BD}} = \frac{2\Sigma_{\mathrm{BD}}}{1+\gamma}  =  \frac{4+2\omega}{3+2\omega} \Sigma_{\mathrm{BD}},
\end{eqnarray}
where $\xi = 0$ leads to $\Sigma_{\mathrm{BD}} = \mu_{\mathrm{BD}} =1$ with $\omega \rightarrow \infty$.
In the PBD case ($\lambda =0$) \cite{Avilez:2013dxa} report $0.981 \leq \Sigma_0 \leq 1.285$ at the 99 per cent confidence level from cosmological tests. The Solar System bound on the time variation of the Newton's gravitational constant is $\vert \dot{G}_{\mathrm{N}}/G_{\mathrm{N}}\vert < 1.3 \times 10^{-12}\,\mathrm{yr}^{-1}$ \cite{Williams:2004qba}, where $G_{\mathrm{N}} = G \mu_{\rm{local}}$. In general, the local value $\mu_{\rm local}$ is determined by the present day value of $\phi$ in our Solar System. Yet, if no screening mechanism is assumed, $\mu_{\rm local}$ may coincide with the background value $\mu_0 = \mu(\phi = \phi_0)$ in the weak-field condition with the quasi-static approximation (see for example \cite{Damour:1992kf, Damour:1993id}).

\section{ model parameters } \label{sec:3}
\begin{center}
\begin{figure}
\includegraphics[width= \columnwidth]{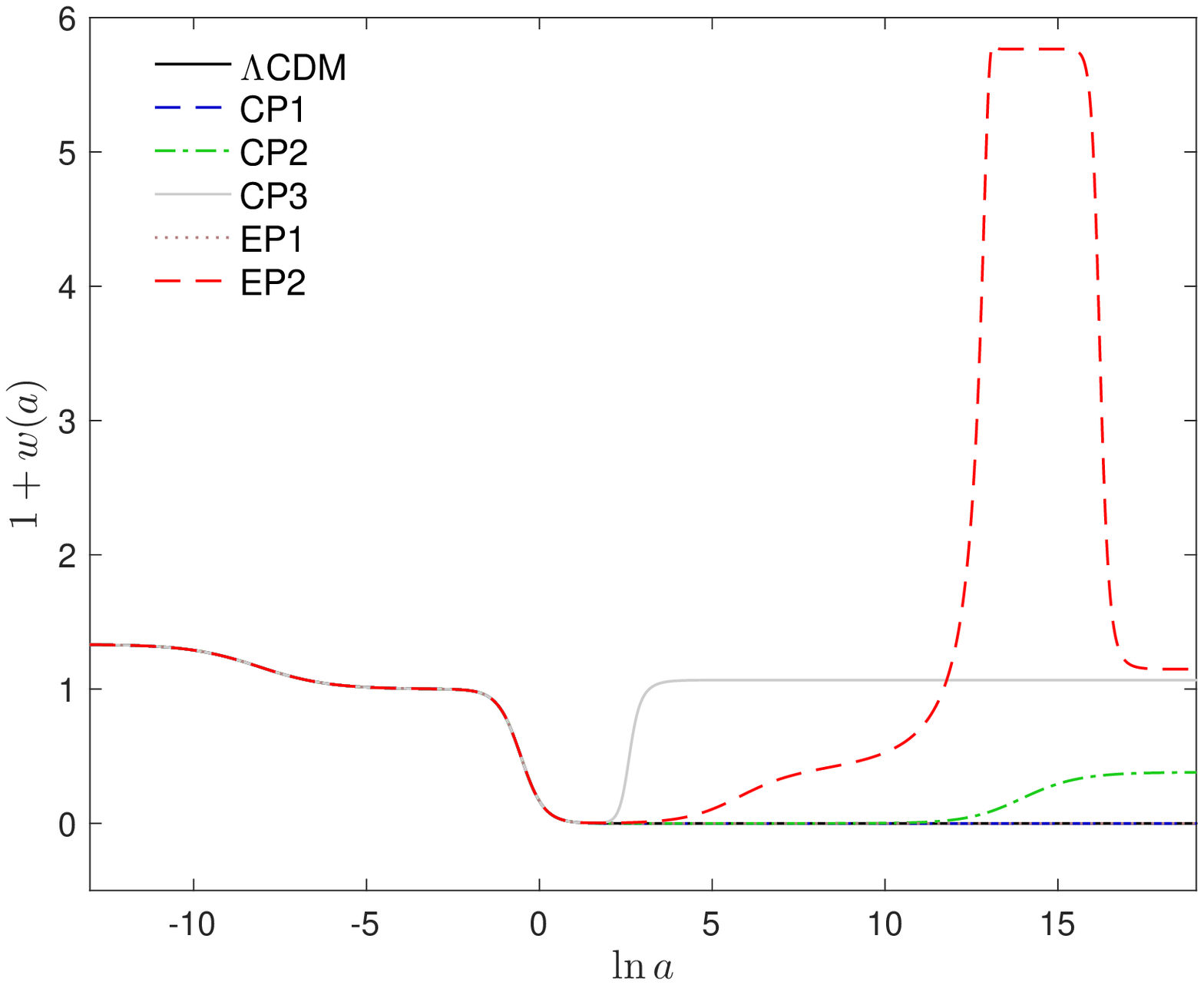}
\includegraphics[width= \columnwidth]{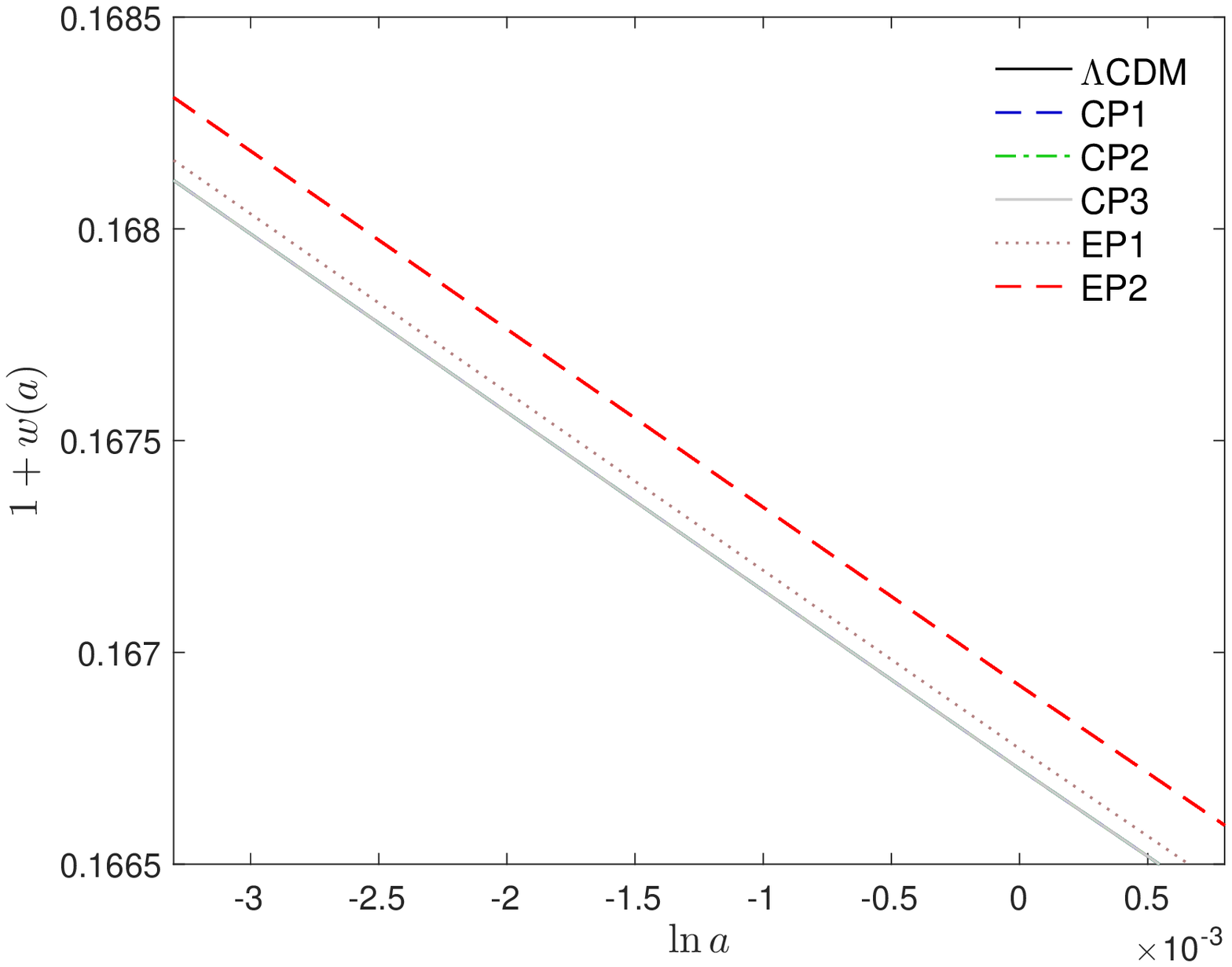}
\caption{Full-time (upper panel) and late-time (lower panel) evolutions of the background equation of state in models with parameters given by: CP1[($\lambda, \xi$) = (0, -0.2)], CP2[($\lambda, \xi$) = (0, 0.2)], CP3[($\lambda, \xi$) = (0, 2)], EP1[($\lambda, \xi$) = (0.02, -0.2)], EP2[($\lambda, \xi$) = (0.02, 0.2)].}
\label{fg:1}
\end{figure}
\end{center}

\begin{center}
\begin{figure}
\includegraphics[width= \columnwidth]{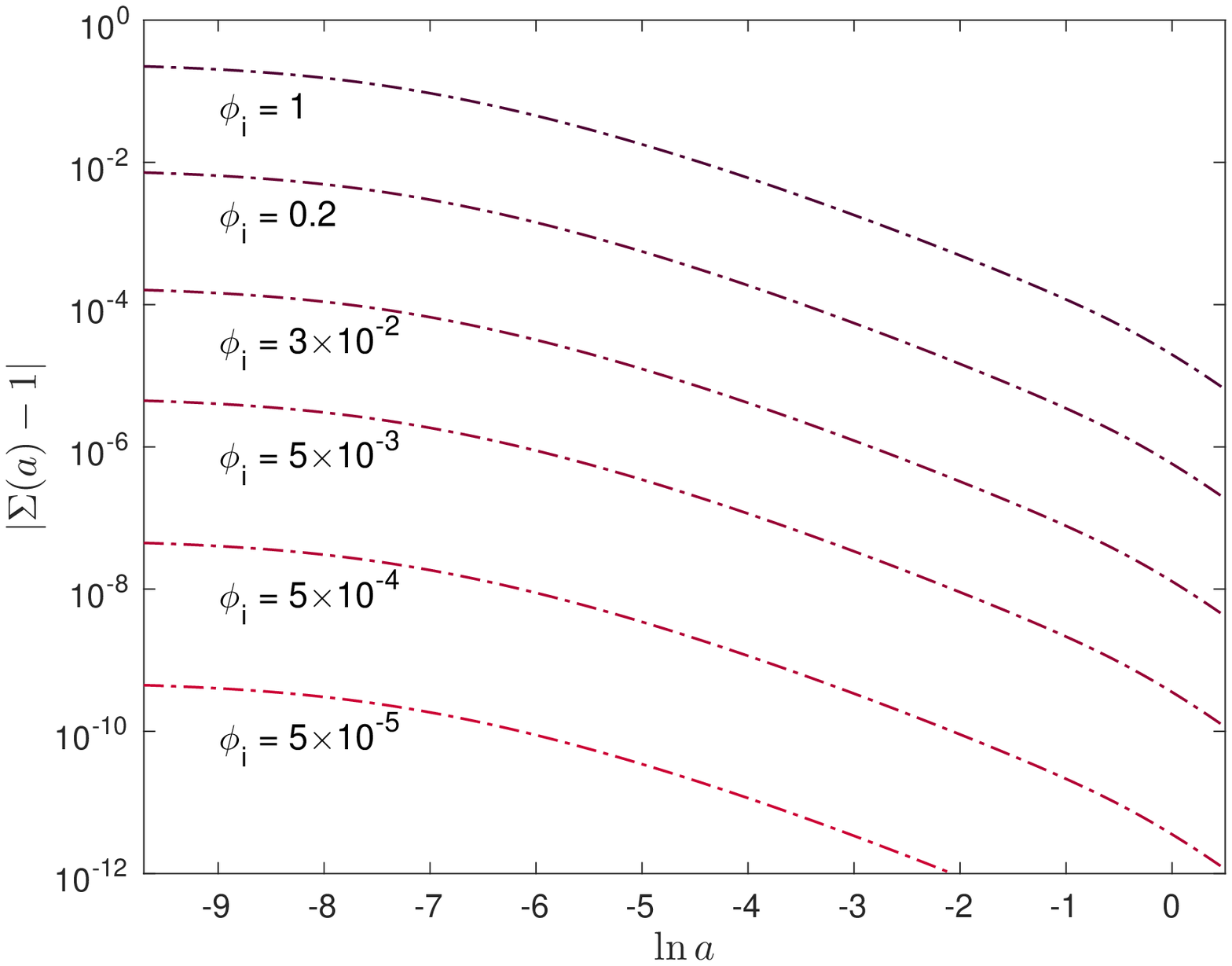}
\includegraphics[width= \columnwidth]{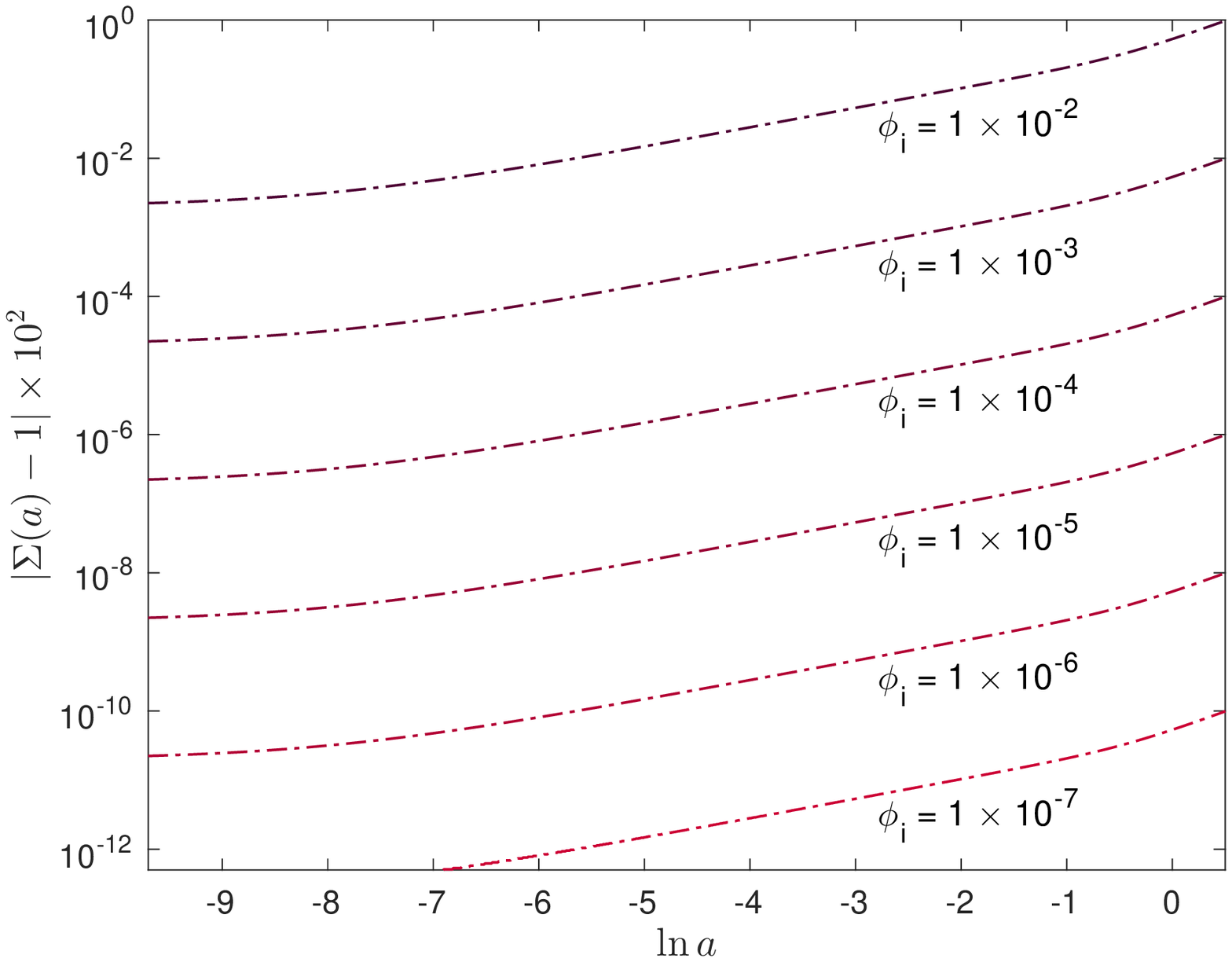}
\caption{Time evolution of $\Sigma$ in the constant potential model ($\lambda = 0$) with the initial condition $\phi^\prime =0$ at $N_i = - 20$, where the non-minimal coupling is $\xi = -0.2$ (upper panel) and $\xi = 0.2$ (lower panel).}
\label{fg:2}
\end{figure}
\end{center}

Here we outline the generic feature of the parameter dependence in the exponential potential model \eqref{eq:V}, and we refer the interested reader to subsections for the analytic solutions upto the asymptotic future. 
In a constant potential case ($\lambda = 0$), the background equation of state $w$ is basically indistinguishable from that of the $\Lambda$CDM model, but $w$ may notably deviate from $-1$ in the near future if $\xi > 0$, see the upper panel of Fig. \ref{fg:1}. In cases where $\lambda > 0$, the actual value of $w$ given by the numerical result in the lower panel of Fig. \ref{fg:1} can differ from that of the $\Lambda$CDM model, and the difference is enhanced if $\xi > 0$. 
Viable cosmic expansions can be realized even if $\xi > 1.5$, but the tracking basin of attraction is tightly restricted.
For $\lambda \geq 0$, the $\Sigma$ function tends to deviate from (converge to) unity since matter became to dominate the universe, given that $\phi$ is govern by an increasing (decreasing) mode when $\xi > 0$ ($\xi < 0$), as shown in Fig. \ref{fg:2}.

\subsection{Constant potential models}
Let us study the case with an absolutely flat potential $\lambda = 0$, where $V(\phi)=V_0$ and $V_\phi=0$. In this case, the Klein-Gordon equation~\eqref{eq:eom_efolding2} is reduced to
\footnote{We assume the background equation of state is a constant with $w=1/3$ during the epoch of radiation domination (RD) and $w=0$ during matter domination (MD).}
\begin{equation}
\label{eq:eom_efolding3}
\phi^{\prime\prime} + \frac{3}{2}( 1 - w ) \phi^{\prime} - 3 \xi \left( 1 - 3 w \right) \phi = 0.
\end{equation}
If $w$ is a constant, this equation can be exactly solved as
\begin{equation}
\label{eq:scaling_solution}
\phi(N)=C_{+} e^{L_{+} N} + C_{-} e^{L_{-} N},
\end{equation} 
where $C_{\pm}$ are constants to be determinated by initial conditions and
\begin{equation}
\label{eq:Lpn}
L_{\pm} (w,\xi)=-\frac{3}{4}(1-w)\pm\sqrt{\frac{9}{16}(1-w)^2+3\xi(1-3w)}.
\end{equation}
For $\xi>0$, one finds $L_{+} > 0$ and thus $e^{L_{+} N} = a^{L_{+}}$ corresponds to an increasing mode, while $e^{L_{-} N} = a^{L_{-}}$ is a decreasing mode as $L_{-} < 0$. For $\xi=0$, $L_{+} = 0$ holds, showing the constant mode of a massless scalar field. For $\xi<0$, no increasing mode exists since $L_{\pm}\leq 0$.
In what follows we examine the evolution of $\phi$ with (a) $\xi \leq 0$, (b) $0< \xi \leq 3/2$ and (c) $\xi > 3/2$, respectively.

\subsubsection*{(a) \; $\xi \leq 0$}

If $\xi = 0$, $\phi$ is governed by a constant mode and the kinetic energy decays rapidly. Eventually, we find $\rho_\phi = V_0$ and $w_\phi = -1$ for arbitrary initial conditions. In this limit the quintessence field reproduces the result of a cosmological constant.

For $\xi < 0$, we have $(L_{+}, L_{-}) = (0, -1)$ in RD, and $(L_{+}, L_{-}) < (0, 0)$ in MD. If $\phi$ is released from rest, the decreasing mode ($\propto e^{L_{-} N}$) becomes negligible with the expansion of the universe and the solution~\eqref{eq:scaling_solution} can be approximated to be
\footnote{Initial conditions with $C_+=0$ and $C_- \neq 0$ imply $\phi= - \phi^\prime$ at the initial time. Since $L_- =-1$ during RD, one finds that $\phi\propto e^{L_- N}\propto 1/a$ and $\rho_\phi\propto H^2\phi^2\propto a^{-6}$, similar to the case in which the initial energy density is dominated by the kinetic energy ($\rho_\phi\approx \dot{\phi}^2/2$).}
\begin{equation}
\label{eq:scaling_solution_RDMD}
\phi = C_{+} e^{L_{+} N}.
\end{equation}
Taking this solution into Eq.~\eqref{eq:rho_phi_ST}, the energy density reads
\begin{equation}
\rho_\phi= V_0 + \frac{1}{\kappa^2}\left( \frac{L_{+}^2}{2} - 3 \xi - 6 \xi L_+ \right) H^2 \phi^2 .
\end{equation}
Assuming that the initial condition satisfies $\rho_\phi\gg V_0$, the energy density is then reduced to
\begin{equation}
\label{eq:rho_phi_V0_minus}
\rho_\phi \approx \rho_\phi^{nc} =  \frac{1}{\kappa^2}\left( \frac{L_+^2}{2} - 3 \xi - 6 \xi L_+ \right) H^2 \phi^2 .
\end{equation}
Here, $\rho_\phi\propto H^2\phi^2\propto a^{2 L_+ - 3 (1+w)}$ indicates that the equation of state takes the form 
\begin{equation}
\label{eq:eos_general}
w_\phi (w,\xi)= w-\frac{2}{3}L_+ (w,\xi),
\end{equation}
resulting in $w_\phi (1/3,\xi)=1/3$ and $w_\phi (0,\xi)=1/2 - \sqrt{1/4+(4\xi)/3}$.

Given that $w_\phi > 0$ in MD when $\xi < 0$, $\rho_\phi$ decays faster than the matter density as $\phi$ approaches zero. Eventually, the value of $\phi$ is frozen around zero and one reproduces the behavior of a cosmological constant with $w_\phi \rightarrow -1$ as $\rho_\phi \rightarrow V_0$.

\subsubsection*{(b) \; $0 < \xi \leq 3/2$}


In this case, the initial density of $\phi$ may be negative. Given that $L_+ > 0$ holds during both RD and MD, $e^{L_+ N}$ is always increasing with time. Assuming that $\phi$ is released from rest, we can adopt the solution~\eqref{eq:scaling_solution_RDMD}, and the evolution essentially has three stages:
 
(i) During RD or MD where $\rho_\phi$ is subdominant, the energy density is increasing with time. To see this, we may denote the solution at this stage as
\begin{equation}
\label{eq:phi_solution_RDMD}
\phi_1 = C_{1} e^{L_{1} N},
\end{equation}
where the decreasing mode has been neglected and $(C_{1},L_{1}) = (C_{+},L_{+})$.

The energy density with $\phi=\phi_1$ takes the form
\begin{equation}
\label{eq:rho_phi_V0_RDMD}
\rho_\phi=  \frac{1}{\kappa^2}\left( \frac{L_1^2}{2} - 3 \xi - 6 \xi L_1 \right) H^2 \phi_1^2 ,
\end{equation}
where it can be checked that $\rho_\phi < 0$ in both RD and MD. One may deduce
$w_\phi = w_1$ from Eq.~\eqref{eq:rho_phi_V0_RDMD} as $w_1(w,\xi) = w-\frac{2}{3}L_1(w,\xi)$, which gives
\begin{equation}
\label{eq:eos_V0_MD}
w_1= \frac{1}{3}\;\;\;(\mathrm{RD})\;\;\;\mathrm{and}\;\;\; w_1 =\frac{1}{2} - \sqrt{\frac{1}{4}+\frac{4}{3}\xi}\;\;\;(\mathrm{MD}),
\end{equation}
where $-1 \leq w_1 < 0$ in MD for $0 < \xi \leq 3/2$.

(ii) Once the density of $V_0$ is dominant, the universe will undergo a de Sitter expansion. Consequently, we can assume that $w = -1$ with $H^\prime = 0$ so that $H = H_{\mathrm{ds}}$ is a constant. Accordingly, Eq.~\eqref{eq:eom_efolding1} is simplified as
\begin{equation}
\label{eq:eom_efolding_dS}
\phi^{\prime\prime} + 3 \phi^\prime - 12 \xi \phi = 0.
\end{equation}
The corresponding solution (keeping only the increasing mode) reads
\begin{equation}
\label{eq:phi2}
\phi_2 = C_{2} e^{L_{2} N},
\end{equation}
where
\begin{equation}
\label{eq:L2}
L_{2}= - \frac{3}{2} + \sqrt{\frac{9}{4} + 12\xi},
\end{equation}
and the energy density~\eqref{eq:rho_phi_ST} is of the form
\begin{equation}
\label{eq:rho_phi_V0_dS}
\rho_\phi= V_0 + \frac{1}{\kappa^2}\left( \frac{L_2^2}{2} - 3 \xi - 6 \xi L_2 \right) H_{\mathrm{ds}}^2 \phi_2^2 .
\end{equation}
Note that $L_2 > 0$ since $\xi > 0$ and therefore, $\phi$ remains increasing with time during the $V_0$ domination epoch.

(iii) Since $\phi_2^2\propto a^{2L_2}$ is increasing with time, the subdominant energy density of $\rho_\phi$ (the second term in the right-hand side of Eq.~\eqref{eq:rho_phi_V0_dS}) eventually becomes comparable with $V_0$, where the de Sitter expansion is interrupted. 
To calculate the cosmological evolution after the $V_0$-domination, we study the asymptotic universe, which has a constant equation of state $w=w_3$, satisfying
\begin{equation}
\label{eq:eom_efolding_f}
\phi^{\prime\prime} + \frac{3}{2}( 1 - w_3 ) \phi^\prime - 3 \xi \left( 1 - 3 w_3 \right) \phi = 0.
\end{equation} 
Again, by neglecting the decreasing mode, the solution is given by
\begin{equation}
\label{eq:phi3}
\phi_3 = C_{3} e^{L_{3} N},
\end{equation}
where
\begin{equation}
\label{eq:L3}
L_{3}= - \frac{3}{4}(1-w_3) + \sqrt{\frac{9}{16}(1-w_3)^2 + 3\xi (1-3 w_3)}.
\end{equation}
Provided that the scalar field is the dominant species of the universe, the Friedmann equation yields
\begin{equation}
\label{eq:Friedmann_f}
3H^2=\kappa^2\rho_\phi,
\end{equation}
where the energy density
\begin{equation}
\label{eq:rho_phi_V0_f}
\rho_\phi= V_0 + \frac{1}{\kappa^2}\left( \frac{L_3^2}{2} - 3 \xi - 6 \xi L_3 \right) H^2 \phi_3^2 .
\end{equation}

For a sufficiently large $N$ such that $\phi_3 \gg 6/(L_3^2 - 6 \xi - 12 \xi L_3)$, the left-hand side of Eq.~\eqref{eq:Friedmann_f} 
is negligible and the lowest order of the Friedmann equation gives $\rho_\phi= V_0+\rho_\phi^{nc}=0$. The asymptotic solution shows a fine cancellation between $V_0$ and $\rho_\phi^{nc}$ (the second term in the right-hand side of Eq.~\eqref{eq:rho_phi_V0_f}). This cancellation implies $\phi_3^2\sim H^{-2}\propto a^{3(1+w_3)}$, which leads to
\begin{equation}
\label{eq:eos_V0_f015}
w_3=\frac{-3+2\xi}{3(1+2\xi)}, \;\;\; \mathrm{and} \;\;\; L_3=\frac{4\xi}{1+2\xi}, \;\;\;(0 < \xi \leq 3/2).
\end{equation}
Given that $-1< w_3 \leq 0$ with $0 < \xi \leq 3/2$, the scalar field will keep overtaking the matter density after the epoch of the de Sitter expansion (the phase of the $V_0$-domination), as seen from Fig.~\ref{fg:1}.

\subsubsection*{(c) \; $\xi > 3/2$}


If the initial value of $\rho_\phi$ is too small such that it remains negative during MD, Eq.~\eqref{eq:eos_V0_MD} shows that $w_1 < -1$ for $\xi > 3/2$ and therefore, the energy density $\rho_\phi$ will always be dominated by the non-minimal coupling term ($\rho_\phi^{nc}$) as given by Eq.~\eqref{eq:rho_phi_V0_RDMD}. In this case, $\rho_\phi$ is always negative, and the constant potential $V_0$ is never important so that there exhibits no epoch of the cosmic acceleration.

On the other hand, if $\rho_\phi$ can be positive before MD, $V_0$ will dominate the universe and drive a de Sitter expansion. Similarly, this de Sitter expansion shall be interrupted with the growths of $\phi$ and $\vert\rho_\phi^{nc}\vert$, where the resulting solution has $w=w_\phi=(-3+2\xi)/[3(1+2\xi)]$. However, since $w_\phi > 0$ for $\xi > 3/2$, the matter density $\rho_M$ may eventually catch up with $\rho_\phi$ and become one of the dominant density component.

Let us assume that the asymptotic solution for $\xi > 3/2$ has a constant equation of state $w_3$ with $\phi_3$ taken the form of Eq.~\eqref{eq:phi3}. For arbitrary initial conditions, the final attractor is a fine cancellation between $\rho_M$ and $\rho_\phi^{nc}$, leading to the lowest order Friedmann equation $\rho_M + \rho_\phi^{nc}=0$. 
Since $\rho_\phi^{nc}\propto H^2\phi_3^2$, the Friedmann equation indicates $\phi_3^2\sim H^{-2}a^{-3}\propto a^{3 w_3}$, which results in
\begin{equation}
\label{eq:eos_V0_xi>1.5}
w_3=\frac{4\xi}{3(1+4\xi)}, \;\;\; \mathrm{and} \;\;\; L_3=\frac{2\xi}{1+4\xi} , \;\;\;(\xi > 3/2).
\end{equation}

The condition of the late-time cosmic acceleration restricts the value of $\phi$ at the initial time. Since $\phi$ is a constant during RD, a non-negative $\rho_\phi$ shall satisfy $3\xi H_{\rm eq}^2 \phi_i^2 \leq \kappa^2V_0 \simeq 3H_0^2$, where $H_{\rm eq}$ is the Hubble parameter at the matter-radiation equality. This condition is approximately $\phi_i^2 \leq \xi^{-1}\times 10^{-17}$ for $\xi > 3/2$, assuming that the equality temperature $T_{\rm eq} \approx 5.5$ eV.

\subsection{ Exponential potential models}

We now revisit the model with $V(\phi)=V_0\, e^{-\lambda \kappa \phi}$, where $\lambda$ is a non-negative constant. Rescaling $\phi$ by $\kappa\phi$, we may rewrite Eq.~\eqref{eq:eom_efolding2} with the exponential potential as
\begin{equation}
\label{eq:eom_exp}
\phi^{\prime\prime} + \frac{3}{2}( 1 - w ) \phi^{\prime} - 3 \xi \left( 1 - 3 w \right) \phi -\lambda \frac{\kappa^2 V_0}{H^2} e^{-\lambda \phi} = 0.
\end{equation}
Analytical solutions in the minimal coupling limit ($\xi=0$) are shown in \cite{Copeland:1997et, Billyard:1998hv}, where attractor solutions of the $\phi$-domination are known to exist for $\lambda^2 < 6$. For $3(1+w) < \lambda^2 < 6$, where $w$ is the background equation of state, the asymptotic attractor is a scaling solution with $\Omega_\phi=3(1+w)/\lambda^2$ and $w_\phi = w$. For $0 < \lambda^2 < 3(1+w)$, the $\phi$-domination solution is stable, and the final state gives $\Omega_\phi=1$ with $w_\phi = -1+\lambda^2/3$. One can see that the cosmic acceleration can be realized in the regime  $0 < \lambda < \sqrt{2}$, where the asymptotic attractor solution conducts $-1 < w_\phi < -1/3$. 

Since $V_0$ corresponds to the energy scale of the dark energy domination, one finds that the potential term is negligibly small in the early time ($N\ll 0$), where the equation of motion~\eqref{eq:eom_exp} coincides with that of the constant potential, as given by Eq.~\eqref{eq:eom_efolding3}. For simplicity, we assume that $\phi$ is released from rest so that the solution~\eqref{eq:scaling_solution_RDMD} can be applied with the evolutions discussed as follows.

\subsubsection*{(a) \; $\xi < 0$}

For $\xi < 0$, the non-minimal coupling term in the Klein-Gordon equation tends to pull back the scalar field from rolling down the potential, and the solution of $\phi$ is a constant mode in RD and a decreasing mode in MD (see Sec. III A). Even if $\phi$ is initially at rest, $\rho_\phi$ can be much greater than the potential energy, and it takes the form of Eq.~\eqref{eq:rho_phi_V0_minus} with a nearly constant equation of state as given by Eq.~\eqref{eq:eos_general}. In this case $\rho_\phi$ is decaying with time for $N\ll 0$, and eventually approaches $V(\phi)$ as $\phi \rightarrow 0$.

Whenever $\rho_\phi = V(\phi)$ is reached, the $\phi$-field will be temporarily frozen by the friction of the Hubble expansion with $w_\phi\simeq -1$ until $\rho_\phi$ starts to dominate the universe.

The asymptotic attractor solution is the stable point balanced between $V(\phi)$ and the non-minimal coupling term. 
Taking $\phi$ to be asymptotically stabilized at some value $\phi_f$ with $H = H_f$ a constant, we may obtain $3H_f^2 = \kappa^2 V_0 e^{-\lambda \phi_f} -3 \xi H_f^2 \phi_f^2$ from the Friedmann equation and $-12\xi H_f^2 \phi_f = \lambda \kappa^2 V_0 e^{-\lambda\phi_f}$ from the Klein-Gordon equation. It is straightforward to find that
\begin{equation}
\label{eq:nxi_exp_f}
\phi_f = -\frac{2}{\lambda} + \sqrt{\frac{4}{\lambda^2}-\frac{1}{\xi}}, \;\;\;
\mathrm{and}\;\;\;
H_f^2 = \frac{\kappa^2 V_0 }{3(1+\xi\phi_f^2)} e^{-\lambda\phi_f},
\end{equation}
where the final energy density reads 
\begin{equation}
\rho_\phi = \frac{V_0}{1+\xi\phi_f^2} e^{-\lambda\phi_f}.
\end{equation}
It can be checked that $\phi_f < 1/\sqrt{-\xi}$ is satisfied for an arbitrary value of $\lambda$, and thus $\rho_\phi$ will not reach the singularity when $1+\xi\phi^2=0$.

\subsubsection*{(b) \; $\xi > 0$}

The non-minimal coupling term with $\xi > 0$ acts as an additional force in the Klein-Gordon equation that pushes $\phi$ to run downhill, and thus $\phi$ is govern by an increasing mode during RD and MD (see Eq.~\eqref{eq:phi_solution_RDMD}). 
Let us schematically divide the evolutions into five stages:

(i)
For $\xi > 0$, a wide range of initial conditions can lead to a negative energy density during RD and MD with $\vert\rho_\phi\vert > V(\phi)$, where $\rho_\phi$ is given by Eq.~\eqref{eq:rho_phi_V0_RDMD} and  increasing toward $V(\phi)$. Once $\rho_\phi = V(\phi)$ is reached, the energy density becomes positive and the value of $\phi$ remains increasing due to the effect of the non-minimal coupling.

(ii)
Once the $\phi$-field dominates the total energy density, the universe experiences a (quasi) de Sitter expansion for a sufficiently small $\lambda$. 
The solution and the energy density of $\phi$ at this stage are given by Eqs.~\eqref{eq:phi2} and \eqref{eq:rho_phi_V0_dS}, respectively.

(iii)
Similar to the case of a constant potential, this de Sitter expansion is going to be interrupted with the increase of $\phi$, once the contribution of the non-minimal coupling term in $\rho_\phi$ is comparable with $V(\phi)$. For $0 < \xi < 3/2$, the solution that follows the de Sitter expansion is a power-law expansion with $w = w_\phi = (-3+2\xi)/[3(1+2\xi)]$ from a temporary fine cancellation of the energy density (that is $\rho_\phi=0$, see Eq.~\eqref{eq:eos_V0_f015}).

(iv)
As $\phi$ keeps growing, $V(\phi)$ is exponentially decaying with time so that at some epoch the potential energy suddenly transits to the kinetic energy, leading to a phase of the super kinetic-energy domination with $w_\phi > 1$. Given that $V(\phi)$ becomes negligible, the lowest order of the Friedmann equation gives
\begin{equation}
\label{eq:FR_exp_4}
0 = \frac{1}{\kappa^2} \left( \frac{L_4^2}{2}-3\xi-6\xi L_4\right) H^2\phi_4^2,
\end{equation}
where $\phi_4 = C_4 e^{L_4 N}$ is the solution of the Klein-Gordon equation with the potential term neglected, and 
$L_4= -3/4(1-w_4)+\sqrt{9/16(1-w_4)^2 + 3\xi (1-3w_4)}$.
Since Eq.~\eqref{eq:FR_exp_4} implies $L_4 = 6\xi+\sqrt{6\xi(1+6\xi)}$, one can find that
\begin{equation}
w_4 = 1 + 8\xi + 4\sqrt{\frac{2\xi}{3}(1+6\xi)}.
\end{equation}

(v)
The density $\rho_\phi$ once again turns into negative when $V(\phi)$ is subsidiary with the increase of $\phi$. We may denote the solution at the final stage as $\phi_5= C_5 e^{L_5 N}$, where $L_5= -3/4(1-w_5)+\sqrt{9/16(1-w_5)^2 + 3\xi (1-3w_5)}$. The asymptotic attractor is a scaling solution in which the density of the $\phi$-field closely tracks that of the dust-like matter, that is $\rho_\phi = -\rho_M \propto a^{-3}$, while the total energy density of the universe $\rho_{\mathrm{tot}}\ll\rho_M$ is always positive. Since $\rho_\phi\propto H^2\phi_5^2 \propto a^{-3}$, we can obtain  
\begin{equation}
\label{eq:solution_exp_f}
w_5=\frac{4\xi}{3(1+4\xi)}, \;\;\; \mathrm{and} \;\;\; L_5=\frac{2\xi}{1+4\xi} , \;\;\;(\xi > 0).
\end{equation}

\section{Observational Constraints} \label{sec:4}

\begin{table}
	\caption{The prior of the cosmological parameters, where parameters in the first column are the baryon density today, the cold dark matter density today, the angular-diameter-distance-to-sound-horizon ratio, the optical depth at reionization, the scalar spectrum index and the log power of the curvature perturbations, while those in the third column are the sum of active neutrino masses, the tensor-to-scalar ratio, current expansion rate ($H_0 = 100 h \rm km\, s^{-1} Mpc^{-1}$), the non-minimal coupling, and the non-minimal-to-total dark energy density fraction today, respectively.}
	\begin{tabular}{ll  ll} 
		\hline
		Parameter & Prior range  & Parameter & Prior range \\ 
		\hline
		$100\Omega_bh^2$ &  $[0.5, 10.0]$ & $\Sigma m_{\nu}  $ &   $[0, 2]$\,(\rm{eV})\\ 
		$100\Omega_ch^2 $ &  $[0.1, 99.0]$ & $ r $ &  $[0, 1]$\\ 
		$100\theta_{\rm MC}$ & $[0.5, 10.0]$ & $H_0$ &  $[20, 100]$\,($ \rm km\, s^{-1} Mpc^{-1}$)\\
		$\tau$ & $[0.01, 0.8]$ & $ \xi $ &  $[0.001, 2]$\\ 
		$ n_s $ &  $[0.9, 1.1]$ & $  \epsilon_0 $ &  $[0, 1]$ \\ 
		$\ln (10^{10} \rm{A_s})$ & $[2.0, 4.0]$ & - & - \\
		\hline
	\end{tabular}
	\label{table1}
\end{table}

In this section we confront the extended quintessence model with cosmological observations.
We define $\epsilon\equiv \Omega_V/\Omega_\phi -1$ to parametrize the fraction of the dark energy density in addition to the potential. The density fraction induced by the non-minimal coupling is then given by $\vert\Omega_{\phi}^{nc}\vert \equiv \vert\epsilon\,\Omega_\phi\vert$, where $\epsilon$ is positive (negative) for $\xi > 0$ ($\xi < 0$).
Given that in the potential \eqref{eq:V} a constant shift of the field value $\phi$ can be absorbed into a redefined $V_0$, we will fix $V_0$ and treat $\phi_0$ as a free parameter to be determined by the present day value $\epsilon = \epsilon_0$. To do so we have to further fix $\lambda$ so that the exponential model has in total 2 + 6 parameters, which are $\xi$ and $\epsilon_0$ plus the six parameters of the standard spatially-flat $\Lambda$CDM model \cite{Ade:2013zuv}. Our optimal choice for the potential is $\lambda = 0$ where the background expansion closely reproduces that of the $\Lambda$CDM model. The prior $\xi > 0 $ is chosen due to the enhancement on the value of $\phi$ at low redshifts, which results in interesting modifications of $w$ and $\Sigma$ to the values in the $\Lambda$CDM model (see Section \ref{sec:3}). On the other hand each sample with $\xi < 0$ eventually converges to the result of $\Lambda$CDM.

We fit the model parameters by virtue of the Markov-chain Monte Carlo approach through the \textsc{cosmomc} program \cite{Lewis:2002ah}. The background and linear perturbation equations given in Section \ref{sec:2} and Appendix \ref{sec:LPE} are numerically solved by the \textsc{mgcamb} package \cite{Lewis:1999bs, Hojjati:2011ix}, where formalisms in both the conformal Newtonian gauge and the synchronous gauge are used. We have included the scalar and tensor mode perturbations in our code both, as considered in \cite{Hwang:1996xh, Zhou:2014fva}. 

We use the selected data sets from the Planck 2015 full-mission release \cite{Adam:2015wua, Aghanim:2015xee} for the CMB temperature and polarizations (and partially for the gravitational lensing \cite{Ade:2015zua}); the 6dF Galaxy Survey \cite{Beutler:2011hx}, the SDSS DR7 \cite{Ross:2014qpa}, the SDSS-III (BOSS) DR10 and 11 \cite{Anderson:2013zyy} for the baryon acoustic oscillations (BAO); the SDSS DR4 and WiggleZ \cite{Blake:2011en} for the matter power spectrum (MPK); the SCP Union 2.1 \cite{Suzuki:2011hu} for the supernova survey; and the CFHTLenS \cite{Heymans:2013fya} for the weak lensing effect.

\begin{center}
	\begin{figure*}
		\includegraphics[width = \linewidth]{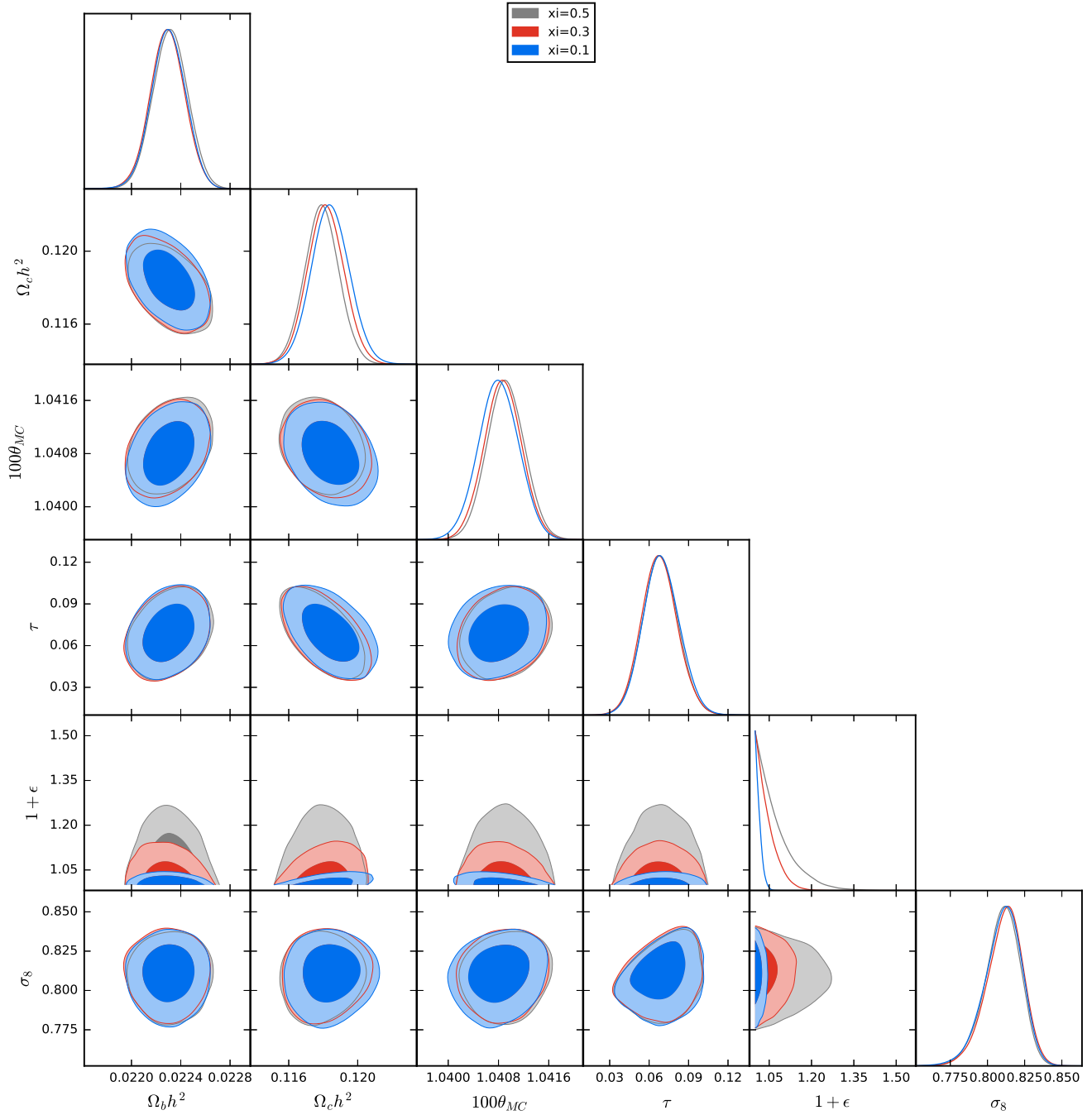}
		\caption{
			Contours of the model parameters in the constant potential model with the fixed non-minimal coupling parameter $\xi$ by using combined PLANCK, BAO, MPK, SNIa and WL data, where the prior of the model parameters is given in Table \ref{table1}.
		}
		\label{fg:fixed_xi}
	\end{figure*}
\end{center}

We first fix the value of $\xi$ to test the allowed region of $\epsilon_0$ by using the combined data of PLANCK + BAO + MPK
with supernova (SNIa) and weak lensing (WL). We find $\epsilon_0 < \{ 0,016, 0.051, 0.085 \}$ and $\epsilon_0 < \{ 0,035, 0.114, 0.201 \}$ at the 68 and 95 per cent confidence levels for $\xi = \{0.1, 0.3, 0.5 \}$, as shown in Table \ref{table_fixed_xi}. Results of the main model parameters with fixed $\xi$ values are given in Fig. \ref{fg:fixed_xi} and Table \ref{table_fix_xi2}. We note that the increase of the $\xi$ value tends to suppress the active neutrino mass sum, as seen by Table \ref{table_fixed_xi}.

\begin{table}
	\caption{Upper bounds of the model parameters $\epsilon_0$ and $\Sigma m_{\nu}$ in the constant potential model with the fixed non-minimal coupling parameter $\xi$ by using the combined PLANCK, BAO, MPK, SNIa and WL data.}
	\label{table_fixed_xi}
	\begin{tabular}{lllllll}
		\hline\hline
		& \multicolumn{2}{l}{$\xi = 0.1$ }  & \multicolumn{2}{l}{$\xi = 0.3$ } & \multicolumn{2}{l}{$\xi = 0.5$ }  \\
		& 68 \% & 95 \%  & 68 \% & 95 \% & 68 \% & 95 \% \\
		\hline
		$\epsilon_0$ & 0.016 & 0.035 & 0.051 & 0.114 & 0.085 & 0.201   \\
		$\Sigma m_{\nu}\,(\rm{eV})\;\;\;\;\; $ & 0.092 & 0.181 & 0.089 & 0.177 & 0.083 & 0.163 \\
		\hline \hline
	\end{tabular}
\end{table}
 
\begin{table}
	\caption{Constraints of the model parameters at the 95 per cent confidence level in the constant potential model with the fixed non-minimal coupling parameter $\xi$ by using the combined PLANCK, BAO, MPK, SNIa and WL data.}
	\label{table_fix_xi2}
	\begin{tabular}{lccc}
		\hline\hline
		& $\xi = 0.1$  & $\xi = 0.3$ & $\xi = 0.5$   \\ 
		\hline
		$100\Omega_bh^2$  & ${2.230}_{-0.026}^{+0.027}$ & ${2.230}\pm 0.027$  &${2.232}\pm 0.027$  \\ [1ex]
		$100\Omega_ch^2$ & ${11.842}^{+0.217}_{-0.213}$ & ${11.815}\pm 0.207$ & ${11.791}^{+0.198}_{-0.197}$   \\[1ex]
		$100\theta_{\rm MC}$ & ${1.0408}\pm 0.0006$ & ${1.0409}\pm 0.0006$ & ${1.0409}_{-0.0005}^{+0.0006}$\\[1ex]
		$\tau$ & ${0.069}^{+0.028}_{-0.026}$ & ${0.068}^{+0.028}_{-0.026}$ & ${0.069}\pm 0.026$   \\[1ex]
		$ n_s $ &  $0.969_{-0.008}^{+0.007}$ & $0.969_{-0.007}^{+0.008}$ & $0.970_{-0.008}^{+0.007}$  \\[1ex]
		$\ln (10^{10} \rm{A_s}) $ & $3.069_{-0.049}^{+0.052}$ & $3.066_{-0.049}^{+0.052}$ & $3.066_{-0.048}^{+0.051}$    \\ [1ex]
		$\Sigma_0$ & $0.993_{-0.011}^{+0.008}$  & $0.986_{-0.022}^{+0.016}$ & $0.983_{-0.032}^{+0.021}$\\
		\hline \hline
	\end{tabular}
\end{table}

\begin{center}
\begin{figure*}
\includegraphics[width = \linewidth]{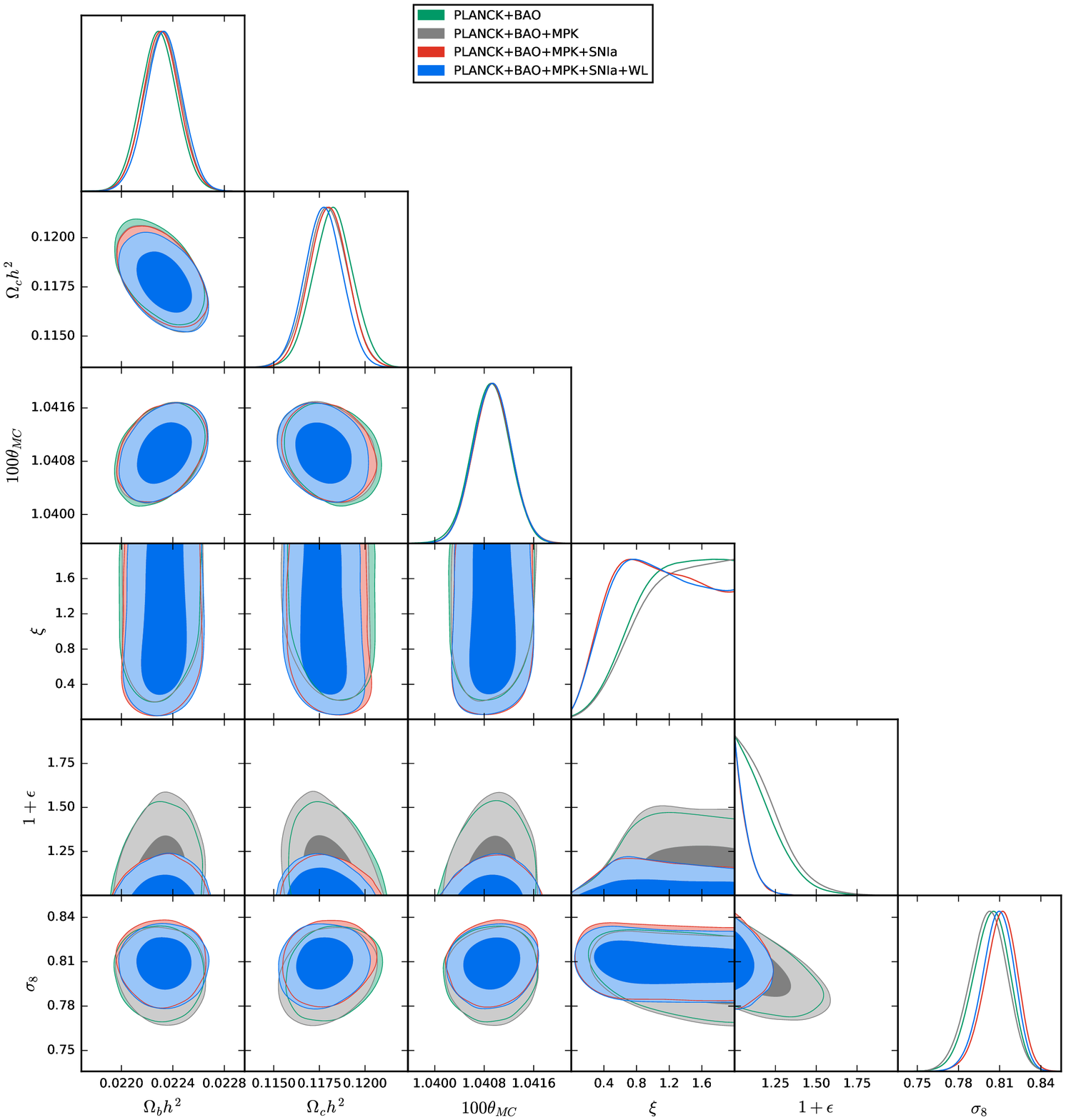}
\caption{
Contour of the model parameters in the constant potential model for the combined PLANCK, BAO, MPK, SNIa and WL data, where the prior of the model parameters is given in Table \ref{table1}.
}
\label{fg:3}
\end{figure*}
\end{center}

We then treat $\xi$ as a free parameter and perform the fitting with respect to various combinations of data sets. 
The best-fit value for $\epsilon_0$ is less than $0.2$ in the first two sets of results based on PLANCK + BAO and their combination with MPK. The best-fit value becomes less than $0.1$ when the SNIa and WL data are included in sequence, as shown in Fig. \ref{fg:3}. In particular, we find $\epsilon_0 < \{ 0.223, 0.456 \}$ at the 68 and 95 per cent confidence levels from the PLANCK + BAO + MPK data, and the constraints are narrow down to $\epsilon_0 < \{ 0.077, 0.180 \}$ by adding SNIa. Given that $\epsilon_0 = 0$ is located inside the $1\sigma$ limit in all four sets of results, we conclude that there is no hint for the existence of the additional density $\Omega_{\phi}^{nc}$, or namely $\Omega_\phi = \Omega_V$ today.

We now discuss the constraint on the non-minimal coupling parameter $\xi$ and the today's value of $\Sigma$, where the full data combinations are listed in Table \ref{tab:fitting}. We observe the result $\xi >  \{1.071, 0.519\}$  at the 68 and 95 per cent confidence levels by using PLANCK + BAO + MPK, but this result can be relaxed to $\xi >  \{0.436, 0.289\}$ at the same confidence levels after combining with the SNIa data. On the other hand, $\Sigma_0 = \{0.974^{+0.025}_{-0.009}, 0.974^{+0.030}_{-0.040}\}$ is found at the 68 and 95 per cent levels with the PLANCK + BAO + MPK data and a stronger constraint $\Sigma_0 = \{0.989^{+0.012}_{-0.004}, 0.989^{+0.014}_{-0.023}\}$ is given with the combination of SNIa.
We can estimate the present day value $\phi_0$ from these results by virtue of the relation
\begin{equation}
\phi_0^2 = \frac{1-\Sigma_0}{\xi\Sigma_0}.
\end{equation}
For example, taking $\Sigma_0 = 0.989$ we find the upper bounds $\phi_0 < \{0.19 , 0.11\}$ for $\xi > \{0.3, 1.0 \}$.

Let us consider more specifically that the extended quint- 
essence model remains a valid approximation on very small scales, such as inside our Solar system. In this case we can use
$\omega_0 > 40000$ as an independent constraint, where Eq. \eqref{eq:omega} leads to
\begin{equation}
\label{BD_constraint}
160000\xi^2 - \xi < \phi_0^{-2}.
\end{equation}
As a result, we have $\phi_0 < \{ 8.3\times 10^{-3}, 2.5\times 10^{-3} \}$ for $\xi = \{0.3, 1.0 \}$.
For models that give a Planck scale with $\phi_0 = 1$, Eq. \eqref{BD_constraint} results in $-0.0025 < \xi < 0.0025$.
Since $1 - \Sigma_0 = \xi\phi_0^2\,\Sigma_0 = 1/(4\xi\omega_0)$, we can use the local constraint $\omega_0 >  40000$ to find that $1 - \Sigma_0 <  \{ 2.1 \times 10^{-5}, 6.3 \times 10^{-6}\}$ for $\xi = \{0.3, 1.0 \}$, 
which are roughly 1000 times stronger than those of the cosmological tests. 

The existence of a positive non-minimal coupling would imply that the future equation of state may change to Eq. \eqref{eq:eos_V0_f015} if $\xi \leq 1.5$ or to Eq. \eqref{eq:eos_V0_xi>1.5} if $\xi > 1.5$. 
Let us define $\phi_\ast$ as the field value at the onset of the $\xi$-induced power-law expansion. Assuming the late-time solution Eq. \eqref{eq:phi2}, we have $\phi_\ast = \phi_0 e^{L_2 N_\ast}$, where $L_2$ is given by Eq \eqref{eq:L2}. Given that the background curvature is decreasing with the increase of $\epsilon$, we may take $3H^2\approx |L_2^2/2-3\xi-6\xi L_2| H^2\phi_\ast^2$ to solve $N_\ast$, which reads
\begin{equation}
N_\ast = \frac{1}{2L_2} \left[ \ln \frac{6}{\left| L_2^2-6\xi-12\xi L_2\right|} - 2 \ln \phi_0\right].
\end{equation}
Setting $\xi = 0.3$, we find $N_\ast > 2.0$ for $\phi_0 < 0.19$ from the cosmological tests and $N_\ast > 5.4$ for $\phi_0 <  8.3\times 10^{-3}$ from the local constraint. Therefore the onset of the $\xi$-induced power-law expansion could occur as early as some $27.6$ ($74.5$) billion years later from the present
for $N_\ast > 2.0$ ($N_\ast > 5.4$).
For the time unit, we have used $H_0 dt= d\ln a =dN$ with $H_0^{-1}\approx 13.8 \times 10^9$ years.



\begin{table*}
 \caption{Constraints of the model parameters $\xi$ and $\epsilon_0$ and the marginalized results of $\Sigma$ in the constant potential models.}
 \label{tab:fitting}
 \begin{tabular}{llllccclll}
  \hline \hline
  & \multicolumn{2}{c}{$\xi$ lower bound}    & \multicolumn{2}{c}{$\epsilon_0$ upper bound}
  & \multicolumn{2}{c}{$\Sigma_0$ }& \multicolumn{2}{c}{$H_0$ ($\rm km\, s^{-1} Mpc^{-1}$)} & Age\, (Gyr) \\
  & 68 \% & 95 \%  & 68 \%  & 95 \%  & 68 \% & 95 \% & 68 \% & 95 \% & 68 \% \\
  \hline
  PLANCK + BAO &  1.043 &  0.503  & 0.200 &  0.415  
  & $0.976^{+0.023}_{-0.008}$ & $0.976^{+0.029}_{-0.038}$ & $66.1^{+1.6}_{-1.0}$ & $66.1^{+2.4}_{-2.7}$ & $13.81\pm 0.03$ \\[1ex]
  PLANCK + BAO + MPK & 1.071 & 0.519  & 0.224 & 0.450  
  & $0.974^{+0.025}_{-0.009}$& $0.974^{+0.030}_{-0.040}$ & $66.0^{+1.6}_{-1.0}$ & $66.0^{+2.4}_{-2.7}$ & $13.81\pm 0.03$\\[1ex]
  PLANCK + BAO + MPK + SNIa & 0.436 & 0.289  & 0.077 & 0.180 
  & $0.989^{+0.012}_{-0.004}$ & $0.989^{+0.014}_{-0.023}$ & $67.3^{+0.9}_{-0.6}$ & $67.3\pm 1.5$ & $13.80\pm 0.03$\\[1ex]
  PLANCK + BAO + MPK + SNIa + WL & 0.794 & 0.300  & 0.076 & 0.183 
  & $0.989^{+0.011}_{-0.003}$ &$0.989^{+0.014}_{-0.023}$ & $67.4^{+0.9}_{-0.6}$ & $67.4^{+1.5}_{-1.4}$ & $13.80^{+0.02}_{-0.04}$\\
  \hline \hline
 \end{tabular}
\end{table*}

\section{Summary and discussions} \label{sec:5}
We have considered one of the simplest extension to the $\Lambda$CDM model based on a quintessence field with 
a non-minimal coupling $\xi$ to gravity. Our principle goal has been to identify the compatibility of such gravitational effect in current observations. 
We have in particular focused on the class of models with very weak restrictions on the today's field value 
$\phi_0$ so that both $\xi$ and $\phi_0$ can be treated as free model parameters.
This condition is suitable for the slow-roll type model $V(\phi) = V_0 e^{-\lambda\kappa\phi}$,
where the potential energy is insensitive to the field value $\phi$ with a sufficiently small $\lambda$. 
In the absolutely flat limit where $\lambda =0$, the theory coincides with the $\alpha = 0$ limit of the tracker model 
$V(\phi) = V_0 (\kappa\phi)^{-\alpha}$ that has a completely vanished tracking basin of attraction.
	However, in general cases the tracker models with $\alpha > 0$ only realize the cosmic acceleration
when $\phi_0\sim 1/\kappa$, provided that $V_0$ is responsible to the energy scale of the dark energy domination~\cite{Zlatev:1998tr,Steinhardt:1999nw,Baccigalupi:2000je}.

We have derived analytic solutions for the background evolution in exponential models, including the possible future time dynamics beyond the de Sitter expansion.
For $\xi < 0$, we have found that the asymptotic attractor is always the de Sitter like solution Eq.~\eqref{eq:nxi_exp_f} 
with the constant Hubble parameter $H_f$, the coupling constant $\xi$, and the final field value $\phi_f$. In 
the special case where $\lambda = 0$, Eq.~\eqref{eq:nxi_exp_f} leads to $\phi_f = 0$ and $H_f=\kappa\sqrt{V_0/3}$. 
For $\xi > 0$, we have shown that the current potential-driven expansion becomes a $\xi$-induced expansion with $w = (-3+2\xi)/[3(1+2\xi)]$. This transition is due to the fine cancellation between the potential energy and the negative density component corresponding to the non-minimal coupling (namely $\rho_\phi^{nc}$). In the $\lambda = 0$ case, the $\xi$-induced power-law expansion can occur even if $\xi > 3/2$, provided that $\rho_\phi$ is positive before the matter domination. However, in this case the solution $w = (-3+2\xi)/[3(1+2\xi)]$ is a saddle point and the final attractor reads $w = 4\xi/[3(1+4\xi)]$. In general cases where $\lambda > 0$, the final attractor can be a scaling solution in which $\vert\rho_\phi\vert$ tracks closely with $\rho_M$. In addition, we have numerically checked that the total density of the universe is always positive even if $\vert\xi\vert$ is very large. 

To probe the existence of $\xi$, we have taken the combined data from type-Ia supernovae (Union 2.1), cosmic microwave background (Planck 2015), baryon acoustic oscillations and matter power spectrum (6dF Galaxy Survey, BOSS and WiggleZ), to the weak lensing (CFHTLenS) measurements. 
With the optimal choice $\lambda =0$ of the model, the background evolution is identical to the fiducial $\Lambda$CDM scenario and the number of the model parameters is the same as that of the PBD research \cite{Avilez:2013dxa}. We have confimed that the most preferable value of $\epsilon_0$ is zero so that there is no hint for the dark energy density additional to that of a pure cosmological constant. This result is true whether we fix a constant value to $\xi$ or treat it as a free parameter.  
Nevertheless, we have demonstrated that the current observations prefer a non-zero value $\xi > 0.289$ for a variable $\xi$, where $\xi = 0$ is excluded outside the 95 per cent confidence region.
	This is in contrast to the extended tracker model~\cite{Baccigalupi:2000je} where a strong constraint $\vert\xi\vert \ll 1$ was found.
Meanwhile, the central value of the effective gravitational constant $\Sigma = G_{\rm{eff}}/G$ today is found to be slightly less than unity as
$\Sigma_0 = 0.989^{+0.014}_{-0.023}$ at 95 per cent level, consistent with the hint of a positive $\xi$.
It is noteworthy that the best-fit sample with a variable $\xi$ parameter brings a slightly better $\chi^2$ results than those with fixed values of $\xi$.
However, a $\xi > 0$ prior always leads to marginally larger $\chi^2$ than the $\Lambda$CDM ($\xi = 0$) case.

	It is noteworthy that the current observations \cite{Ade:2015xua} (see also \cite{Spergel:2006hy}) on the dark energy equation of state
indicate a nearly $2\sigma$ preference in the phantom domain, i.e. $w_{\rm DE} < -1$. This result implies the importance to probe the 
constraint on $\xi$ with a slightly phantom-type background expansion, as a straightforward extension to our current study.
Note that the non-minimal theory given by \eqref{eq:action_STmodel} cannot cross the phantom divide due to its conformal equivalence
to the coupled equintessence model~\cite{Pettorino:2008ez}. 
In addition, it has been found that the gravitational non-minimal coupling $\xi$ can act as a cosmological attractor for
not only the early-time~\cite{Kallosh:2013yoa, Kallosh:2013hoa, Ferrara:2013rsa, Galante:2014ifa, Odintsov:2016jwr, Odintsov:2016vzz} 
but also for the late-time~\cite{Linder:2015qxa} cosmic accelerations. In this class of models one may go beyond the slow-roll conditions for $V(\phi)$,
and obtain an important density fraction $\vert\Omega_\phi^{nc}/\Omega_V\vert \sim \mathcal{O}(1)$ today. 

	We also emphasize that the working frame used in this study is based on the matter point of view 
	\cite{Domenech:2015qoa,Domenech:2016yxd} (see also \cite{Bahamonde:2016wmz,Bahamonde:2017kbs}) 
	in which both matter and radiation are minimally coupled with gravity and the
dark energy field. This physical frame appears to be the most natural choice when performing the fitting with 
the cosmological priors given in Table~\ref{table1}, whereas the possible distinction on observables in different conformal frames
is an issue goes beyond the scope of the current paper.

Finally, we remark that if a future measurement would confirm $1 - \Sigma_0 > 0$, it can support the exsitence of a positive coupling $\xi$ but will falsify at the same time the contribution of many types of screening mechanisms near the Solar System \cite{Pogosian:2016pwr}.
On the other hand, a future measurement with $1 - \Sigma_0 < 0$ (as already hinted in \cite{Ade:2015rim}) will not in favor of a negative $\xi$ but instead can rule out the whole class of massless extended quintessence models, given that $\xi < 0$ only leads to the decay of the field value so that $\Sigma$ always converges to the $\Lambda$CDM limit where $\Sigma_0 = 1$.

\section*{Acknowledgments}
We are grateful to Shinji Tsujikawa for helpful comments.
The work was supported in part by National Center for Theoretical Sciences, 
MoST (MoST-104-2112-M-007-003-MY3), National Science
Council (NSC-101-2112-M-007-006-MY3) and National Tsing Hua
University (104N2724E1).





\appendix
\section{Linear perturbation equations}
\label{sec:LPE}

We show the linear perturbation equations of the model \eqref{eq:action_STmodel} with metric in the conformal Newtonian gauge of the form
\begin{eqnarray}
\label{eq:newtonian_gauge}
ds^2 = a^2(\tau) \left[ -\left( 1+2\Psi \right) d\tau^2 + \left( (1 - 2 \Phi) \delta_{ij} + D_{ij} \right) dx^i dx^j \right],
\end{eqnarray}
where $\tau$ is the conformal time, $\Psi$ and $\Phi$ are respectively the Newtonian and the curvature potentials, and $D_{ij}$ is a symmetric tensor that satisfies the traceless and transverse conditions: $D_{ii} =0$, $\partial_i D_{ij} =0$.
The components of the perturbed energy-momentum tensor $T_{\mu}^{\;\nu}$ are 
$T_0^{\;0}= -\left( \rho_M + \delta \rho_M \right)$, $T_i^{\;0} = \left( \rho_M + P_M \right) \partial_i v_{M}$,
$T_i^{\;j} = \left( P_M + \delta P_M \right) \delta_i^{\;j} +\pi_i^{\;j}$,
where $v_M$ is the peculiar velocity of matter, $\pi^i_j$ is the traceless anisotropic stress, and $i,j=1,2,3$.
It is convenient to define the comoving matter density perturbation $\Delta$ such that
\begin{equation}
\rho_M \Delta = \delta \rho_M + 3 \frac{\mathcal{H}}{k}\left( \rho_M +P_M\right) v_M,
\end{equation}
where $\mathcal{H} = da/d\tau = a H$.
 
We decompose the scalar field into a homogeneous part and a perturbation one as $\phi(x)=\phi(\tau)+\delta\phi(\vec{x})$. Let us put the perturbed metric, the perturbed energy-momentum tensor and the perturbed $\phi$ in Eq. \eqref{eq:field_STmodel} to obtain the perturbed field equations as ($\kappa^2_{\rm eff} = \Sigma\kappa^2$)
\begin{eqnarray}
\label{eq:pert-00}
& 2\left[ 3 \mathcal{H} \left( \Phi^{\prime} + \mathcal{H} \Psi \right) -  \partial^2 \Phi \right]  = 
\kappa^2_{\rm eff}(-a^2  \delta \rho_M +  \delta \Theta _{0}^{\; 0}), 
\\ \label{eq:pert-0i}
& 2 \partial^i\left(\Phi^{\prime} + \mathcal{H} \Psi \right) =
\kappa^2_{\rm eff}\left[- a^2  \left( \rho_M + P_M\right) \partial^i v_M +  \delta\Theta _{0}^{\; i}\right],
\end{eqnarray}
where in this section primes are derivatives with respect to $\tau$, and
\begin{eqnarray} \nonumber
\delta \Theta _{0}^{\; 0}  = & \phi^{\prime 2} \Psi - \phi^{\prime} \delta \phi^{\prime}  -  [a^2 V_\phi - 6 \xi\mathcal{H}(\phi^{\prime}+  \mathcal{H} \phi)] \delta \phi   \\
& - 6 \xi  \phi \phi^{\prime}( 2 \mathcal{H} \Psi + \Phi^{\prime}) -  2 \xi \phi (\partial^2 \delta\phi - 3 \mathcal{H} \delta\phi^{\prime}), \\
\delta\Theta _{0}^{\; i} = & (1 + 2\xi) \phi^{\prime} \partial^i \delta \phi - 2 \xi \phi \partial^i \left( \mathcal{H} \delta\Phi - \delta\phi^{\prime} + \phi^{\prime} \Psi \right),
\end{eqnarray}
where $\partial^2 \equiv \delta^{ij}\partial_i\partial_j$. The spatial part is divided into  
\begin{equation}\label{eq:pert-ijsn}
\Phi = \Psi +2 \kappa^2_{\rm{eff}} \xi \phi \delta\phi,
\end{equation}
for the $i \neq j $ components and
\begin{eqnarray}\label{eq:pert-ijse}
 2 &  \left[ \left( \mathcal{H}^2  +   2 \mathcal{H}^{\prime} \right) \Psi + \mathcal{H} \Psi^{\prime} + \Phi^{\prime\prime} + 2 \mathcal{H} \Phi^{\prime} \right] \delta_i^{\; j}  \nonumber\\
& =  \kappa^2_{\rm{eff}} (a^2  \delta P_M \delta_i^{\;j} + \delta\Theta _{i}^{\; j}), 
\end{eqnarray}
for the $i = j$ components, where
\begin{eqnarray}
\delta\Theta _{i}^{\; j}
& = - a^2 \delta_i^{\;j}  V_\phi \delta\phi 
+ (1+4\xi) \delta_i^{\;j} \left( \phi^{\prime} \delta\phi^{\prime} - \phi^{\prime 2} \Psi \right) 
\\\nonumber
& -2 \xi \delta_i^{\;j} \left[ \mathcal{H} \phi^{\prime} - \phi^{\prime\prime} -(\mathcal{H}^2 + \mathcal{H}^{\prime})\phi \right] \delta\phi 
\\\nonumber
& -2 \xi \phi \delta_i^{\;j} \left(2 \mathcal{H} \phi^{\prime} \Psi + 2 \phi^{\prime} \Phi^{\prime} + \phi^{\prime} \Psi^{\prime} + 2 \phi^{\prime\prime} \Psi -\mathcal{H} \delta\phi^{\prime} - \delta\phi^{\prime\prime} \right).
\end{eqnarray}
Similarly, the perturbed Klein-Gordon equation is derived as
\begin{eqnarray}
&  \delta\phi^{\prime\prime} + 2 \mathcal{H} \delta\phi^{\prime} - 6 \xi ( \mathcal{H}^{\prime} + \mathcal{H}^2 ) \delta\phi - \partial^2 \delta\phi + a^2 V_{\phi\phi} \delta\phi \nonumber \\
& + 2 \xi \phi [ 6 ( \mathcal{H}^{\prime} + \mathcal{H}^2 ) \Psi + 9 \mathcal{H} \Phi^{\prime} + 3 \mathcal{H} \Psi^{\prime} + 3 \Phi^{\prime\prime} + \partial^2 (\Psi - 2 \Phi) ] \nonumber \\
& - 4 \mathcal{H} \phi^{\prime} \Psi - 3 \phi^{\prime} \Phi^{\prime} - \phi^{\prime} \Psi^{\prime} - 2 \phi^{\prime\prime} \Psi  = 0.
\end{eqnarray}
The tensor mode perturbation given by the spatial part of the modified Einstein equation \eqref{eq:field_STmodel} is simply
\begin{equation}
\label{eq:dij_ST}
D^{i \prime \prime}_{\;j} + 2 \left( \mathcal{H}+ \xi \Sigma \phi \phi^{\prime}  \right) D^{i \prime}_{\;j} -  \partial^2 D^i_{\;j} = \kappa^2 a^2 \Sigma \pi^i_{\;j} ,
\end{equation}
where $\pi^i_{\;j}$ is assumed to be vanished in this work.

\end{document}